\documentclass[aps, pra, singlecolumn]{revtex4}
\usepackage{amsmath}
\usepackage{dsfont}
\usepackage{graphicx}
\usepackage{graphicx,amsmath,amssymb}

\newcommand{\abs}[1]{\lvert#1\rvert}

\newcommand{\KK}{\mathbf{K}}
\newcommand{\xx}{\mathbf{x}}
\newcommand{\kk}{\mathbf{k}}
\newcommand{\qq}{\mathbf{q}}
\newcommand{\ii}{\mathrm{i}}

\begin{document}

\title{A smooth polaron-molecule crossover in a Fermi system}

\author{ D M Edwards}

\address{Department of Mathematics, Imperial College London, London SW7~2BZ, United Kingdom}


\begin{abstract}The problem of a single down spin particle interacting with
a Fermi sea of up spin particles is of current interest in the field of cold
atoms. The Hubbard model, appropriate to atoms in an optical lattice potential,
is considered in parallel with a gas model. As the strength of an attractive
short-range interaction is increased there is a crossover from ``polaron''
behaviour, in which the Fermi sea is weakly perturbed, to "molecule" behaviour
in which the down spin particle is bound to a single up spin particle. It
is shown that this is a smooth crossover, not a sharp transition as claimed
by many authors. 
 
\end{abstract}

\maketitle

\section{Introduction}\label{sec:intro}

The behaviour of a system in which a mobile particle couples to a fermion
bath has been of interest for at least fifty years. A useful model of such
a system is provided by a Hubbard model~\cite{Hu63} with a single down spin particle interacting with a Fermi sea of up spin particles. The on-site interaction
$U$ may be repulsive ($U>0$) or attractive ($U<0$). For low particle density
the Hubbard lattice model maps onto a continuum gas model with short range
interactions and in this regime the theory of the two models with a single
reversed spin can proceed in parallel. McGuire~\cite{McG65, McG66} found
exact solutions for the gas case in one dimension (1D) with both repulsive
and attractive interactions. This pioneering work was a fore-runner of Lieb and Wu's exact Bethe ansatz solution of the 1D Hubbard model with arbitrary up and down spin densities~\cite{LW69}. Edwards~\cite{Ed90} wrote McGuire's
exact wave function in a simple form which enabled Castella and Zotos~\cite{CZ93}
to calculate spectral properties. It is remarkable that the highly simplified
case of one down spin particle already contains the basic non-Fermi-liquid
features of 1D conductors; in particular the down spin quasiparticle weight
vanishes in the thermodynamic limit.

The repulsive case in higher dimensions with one spin reversal was first
studied in connection with the stability of ferromagnetism in the Hubbard
model~\cite{Na66, Ro69}. In 1D the ferromagnetic state is never stable~\cite{LW69}.
The use of increasingly sophisticated variational wave functions~\cite{vdLE91,Wu96}
has provided strong evidence for the existence of a small region of the phase
diagram (particle density $\rho$ versus $U/t$ where $t$ is the hopping parameter)
of the Hubbard square lattice where a state of complete spin alignment is
stable against reversing a spin.

The discovery of high temperature superconductivity in Cu$\mathrm{0_2}$ planes
launched a new wave of interest in the 2D Hubbard model. The unusual normal
state in these materials led Anderson~\cite{An90} to propose that non-Fermi-liquid
behaviour, similar to that in 1D, might occur in the Hubbard square lattice.
Since in 1D such behaviour is found already in states with a single spin
reversal Sorella~\cite{So94} made accurate calculations of the down spin
quasiparticle weight $Z$ for this case in 2D. Except for the case of a half-filled
up spin band, where the Fermi level is at a van Hove singularity, he found
no evidence of a departure from Fermi liquid behaviour. Z is found not to
fall below 0.1 even for the strongly attractive case $U=-\infty$ with $\rho=1/4$ where the down spin particle might be expected to lose its fermionic character in forming a molecule with an up spin particle.

The latest renaissance of the Hubbard model is associated with the experimental
study of fermionic cold atoms in optical lattices. The related continuum
gas model corresponds to harmonically trapped atoms where in theoretical work a uniform gas is often assumed. Strongly spin polarized Fermi gases can be realised experimentally and there is an excellent review of the experimental
and theoretical situation by Chevy and Mora~\cite{CM10}. They stress the importance
of understanding the case of a single down spin particle coupled to the up
spin Fermi sea by an attractive interaction. For weak interaction the down
spin particle merely disturbs the Fermi sea by creating a few electron-hole
pairs and may be described as a "polaron". For very strong interaction the
down spin may bind to a single up spin to form a "molecule". A number of authors claim that as the interaction strength is increased there is a sharp
polaron-molecule transition with zero down-spin quasiparticle weight on the
molecule side of the transition~\cite{PSa08,PSb08,PDZ09,BM10,TC12,Pa11,PL13}.
The first five of these references are concerned with the 3D system. The last two consider the 2D case and their conclusions conflict with the work
of Sorella discussed in the previous paragraph.

In this paper we argue that for the Hubbard and gas models, in both 2D and 3D, there is a smooth crossover between the polaron and molecule limits. 
   
\section{The exact wave-function for one reversed spin}\label{sec:exact wave-function}

We consider the Hamiltonian
\begin{equation}\label{ham}
H=\sum_{\kk\sigma}\epsilon_{\kk}c_{\kk\sigma}^{\dagger}c_{\kk\sigma}+\frac{g}{V}\sum_{\kk,\kk^\prime,\qq}c_{\kk\uparrow}^{\dagger}c_{\kk^\prime\downarrow}^{
\dagger}c_{\kk^\prime+\qq\downarrow}c_{\kk-\qq\uparrow}
\end{equation}
where $c_{\kk\sigma}$ destroysdestroys a particle with wave-vector $\kk$ and spin
$\sigma$, $g/V$ is the interaction between particles of opposite spin and $V$
is the volume of the system. In the Hubbard case 
\begin{equation}\label{epsilonk}
\epsilon_{\kk}=-t\sum_{\mathbf{R}} {\rm e}^{\ii\kk\cdot\mathbf{R}}
\end{equation}
with summation over nearest neighbours $\mathbf{R}$. Also $g=Ua^d$ where $d$
is the dimensionality and $a=\abs{\mathbf{R}}$ is the lattice constant for
the 1D, square and simple cubic lattices. In the gas case $\epsilon_{\kk}=\hbar^2k^2/2m$,
which corresponds to the long-wavelength limit of~\eqref{epsilonk} with
$\hbar^2/2m=a^2t.$ The wave-vector summations are taken over the Brillouin zone in the Hubbard case and restricted by a suitable cutoff in the
gas case (see e.g.~\cite{ZBP11}).
We are concerned with the case of $N_{\uparrow} \uparrow$ spins and one $\downarrow$
spin. The form of exact wave-function for this case is~\cite{vdLE91}
\begin{equation}\label{psiK}
\vert\psi(\KK)\rangle =\sum_{Q} C(Q)c_{\KK-\mathbf{Q}\downarrow}^{\dagger}\vert
Q\rangle
\end{equation}
where the summation is over all possible $N_{\uparrow}$-tupels $Q$ of wave-vectors
$\kk$ for the $\uparrow$ spin particles. Here
\begin{equation}\label{Q}
\vert Q\rangle=\prod_{\kk\in Q}c_{\kk\uparrow}^{\dagger}\vert 0\rangle \qquad
\mathbf{Q} =\sum_{\kk \in Q}\kk
\end{equation}
where $ \vert 0\rangle$ is the vacuum state. The number of $N_{\uparrow}$-tupels
is clearly $ \binom{N}{N_{\uparrow}}$ where $N$ is the number of wave-vectors
in the zone, equal to the number of lattice sites.

The exact wave-function may be written in an alternative form which emphasizes
the polaronic nature of the $\downarrow$ spin particle. This was done by
Combescot and Giraud~\cite{CG08} for the gas case. The Hubbard case is essentially
the same for $N_{\uparrow}<N/2$ and also for $N_{\uparrow}>N/2$ if we consider
a gas of holes.
We select a set of $N_{\uparrow}$ wave-vectors to form a $\uparrow$ spin
Fermi sea (FS). These will normally lie within the Fermi surface of the non-interacting
system. By using FS as a new vacuum state we can express the exact wave function
of Eq.~\eqref{psiK} in terms of excitation of particle-hole pairs. The Fermi
sea of $N_{\uparrow}\uparrow$ spin particles is given by
\begin{equation}\label{FS}
\vert \rm FS ( N_{\uparrow})\rangle=\prod_{\kk\in  \rm FS (
N_{\uparrow})}c_{\kk\uparrow}^{\dagger}\vert
0\rangle
\end{equation}
and we can rewrite~\eqref{psiK} as
\begin{equation}\label{psipolaron}
\begin{split}
\vert\psi(\KK)\rangle =[\alpha_0 c_{\KK \downarrow}^{\dagger}+\sum_{\kk,\qq}
\alpha_{\kk\qq}c_{\KK -\kk+\qq \downarrow}^{\dagger}c_{\kk\uparrow}^{\dagger}c_{\qq\uparrow}+\dotsb\\
+\frac{1}{(n!)^2}\sum_{(\kk_i)(\qq_j)}\alpha_{(\kk_i)(\qq_j)}c_{\mathbf{P}\downarrow}^{\dagger}\displaystyle\prod_{1\leq
i\leq n}c_{\kk_i\uparrow}^{\dagger}\displaystyle\prod_{1\leq j\leq n}c_{\qq_j\uparrow}+\dotsb]\vert\rm FS ( N_{\uparrow})\rangle
\end{split}
\end{equation}
where states $\kk,\kk_i$ are outside the FS and $\qq,\qq_j$ are inside .The
total wave-vector $\KK$ is relative to the total wave-vector of the assumed
FS. In the general term, with $n$ particle-hole pairs excited, the wave-vector
$\mathbf{P}$ in the $c_{\mathbf{P}\downarrow}^{\dagger}$ operator is given
by
\begin{equation}\label{P}
\mathbf{P}=\KK -\sum_{1\le i\le n}\kk_i+\sum_{1\le j\le n}\qq_j.
\end{equation}
The coefficients $\alpha_{(\kk_i)(\qq_j)}$ are antisymmetric with respect
to the exchange of any of their arguments $\kk_i$ or $\qq_j$. In the Hubbard
case the number of states with n particle-hole pairs is $ \binom{N_{\uparrow}}{n}\binom{N-N_{\uparrow}}{n}
$ and the  last term of the series corresponds to $n=N_{\uparrow}$. Since
\begin{equation}\label{binomial}
\sum_{0\le n\le N_{\uparrow}} \binom{N_{\uparrow}}{n}\binom{N-N_{\uparrow}}{n}=\binom{N}{N_{\uparrow}}
\end{equation}
the total number of independent $\alpha$ coefficients in~\eqref{psipolaron}
is equal to the number of coefficients $C(Q)$ in~\eqref{psiK} as it should
be. Approximate wave-functions obtained by truncating the series~\eqref{psipolaron}
at a small number of terms have been used as variational forms by many authors
in this field, commencing with Chevy~\cite{C06}. We follow Parish and Levinson~\cite{PL13} in denoting the wave-function
which ends with the term involving $n$ particle-hole pairs by $\vert P_{2n+1}(\mathbf{K})\rangle$.

Several authors~\cite{PDZ09,ZBP11,Pa11,PL13} have introduced another set of states which emphasize
the molecular aspect of the problem. Again following the notation of~\cite{PL13}
we write
\begin{equation}\label{M}
\begin{split}
\vert M_{2n}(\KK)\rangle =[\displaystyle\sum_{\kk}\xi_{\kk}c_{\KK-\kk\downarrow}^{\dagger}c_{\kk\uparrow}^{\dagger}
+\displaystyle\sum_{\kk,\kk_1,\qq_1}
\xi_{\kk\kk_1\qq_1}c_{\KK-\kk-\kk_1+\qq_1 \downarrow}^{\dagger}c_{\kk\uparrow}^{\dagger}
c_{\kk_1\uparrow}^{\dagger}c_{\qq_1\uparrow}+\dotsb\\ 
+\frac{1}{(n!)^2}\displaystyle\sum_{\kk(\kk_i)(\qq_j)}\xi_{\kk(\kk_i)(\qq_j)}
c_{\mathbf{P}^{\prime}\downarrow}^{\dagger}c_{\kk\uparrow}^{\dagger}\displaystyle\prod_{1\leq
i\leq n}c_{\kk_i\uparrow}^{\dagger}\displaystyle\prod_{1\leq j\leq n}c_{\qq_j\uparrow}]\vert\rm FS ( N_{\uparrow}-1)\rangle
\end{split}
\end{equation}
where $\vert\rm FS ( N_{\uparrow}-1)\rangle$ represents a FS containing 
$ N_{\uparrow}-1$ states and $\KK$ is the total wave-vector relative to the
total wave-vector of this FS. Here
$\mathbf{P^{\prime}}=\KK-\kk-\sum{\kk_i}+\sum{\qq_j}$. If, as a special case, we take $\xi_{\kk}=\alpha_{0}\delta_{\kk\kk^{\prime}}$,
$\xi_{\kk\kk_i\qq_j}=\alpha_{\kk_i\qq_j}\delta_{\kk\kk^{\prime}}$ and put
 $\vert\rm FS(N_{\uparrow})\rangle=c_{\kk^{\prime}\uparrow}^{\dagger}\vert\rm FS ( N_{\uparrow}-1)\rangle$, we see that $\vert M_{2n}(\mathbf{K})\rangle$ becomes
identical to $\vert P_{2n-1}(\KK-\kk^{\prime})\rangle$. We can include total
wave-vector as a variational parameter and hence define $E_{2n}$ as the lowest
energy, calculated variationally, for  $\vert M_{2n}(\mathbf{K})\rangle$
and $E_{2n-1}$ as the lowest energy for  $\vert P_{2n-1}(\KK)\rangle$.
All energies are for a given particle number and interaction strength. We
have shown that every $\vert P_{2n-1}\rangle$ state is a special case of
a $\vert M_{2n}\rangle$ state and hence its energy is greater than that of this $\vert M_{2n}\rangle$ state and hence greater than $E_{2n}$, the lowest  $\vert M_{2n}\rangle$ energy. Hence $E_{2n-1}\geq E_{2n}$. Similarly we
 can show that every  $\vert M_{2n}\rangle$ state is a special case of a $\vert P_{2n+1}\rangle$ state so that $E_{2n}\geq E_{2n+1}$. Hence

\begin{equation}\label{inequality}
 E_{2n-1}\geq E_{2n}\geq E_{2n+1}.
 \end{equation}
 Parish and Levinson~\cite{PL13} have obtained a similar result, but
 they do not draw the following conclusions:
 
 (i) Curves of $E_{2n-1}$ and $E_{2n}$ as functions of interaction strength,
 for a given particle number, should never cross. Apparent crossings found
 in 3D~\cite{PSa08,PSb08,PDZ09,BM10,TC12} and 2D~\cite{Pa11,PL13} have been interpreted as sharp transitions
 between a "polaron" state $\vert P\rangle$ and a "molecule" state
  $\vert M\rangle$. Sharp transitions of this type should not occur.
  
  (ii) Since every  $\vert M\rangle$ state is a special case of a $\vert P\rangle$ state there is really nothing to be gained by introducing $\vert M\rangle$ states. Combescot et al~\cite{CRLC07} have shown that the simple
ansatz $\vert P_3(\KK)\rangle$ describes a smooth crossover from "polaron" to
"molecule" behaviour quite accurately. The accuracy would be improved using
$\vert P_{5}\rangle$, $\vert P_{7}\rangle$ ... although the numerical calculations
rapidly become unmanageable~\cite{Ta75}. An alternative approach, based on a variational approach which is exact in 1D but containing hardly more
variational parameters than in $\vert P_3(\KK)\rangle$, is described in the
next section.
A variational treatment of the approximate ansatz $\vert P_{2n+1}(\KK)\rangle$
with up to $n$ electron-hole pairs excited leads to an equation for the $\downarrow$
spin excitation energy $\omega$ of the form~\cite{Ta75}
\begin{equation}\label{dyson}
\omega-\epsilon_{\KK}-\Sigma_{2n+1}(\KK,\omega)=0.
\end{equation}
The total energy of the state  $\vert P_{2n+1}(\KK)\rangle$ is $E_0 +\omega$
where $E_0 =\sum\epsilon_{\KK}$ is the energy of the FS state given by Eq.~\eqref{FS}.
The self-energy $\Sigma_{2n+1}(\KK,\omega)$ corresponds diagrammatically
to summing all diagrams with up to $2n+1$ particle lines and all possible
interaction lines linking the $\downarrow$ spin particle line with one of
the $\uparrow$ spin particle(hole) lines. It is exact for the case $N_{\uparrow}=n$,
for all Hubbard parameters $t$ and $U$, and hence gives the correct atomic
limit ($t=0$):
\begin{equation}\label{atomiclimit}
\Sigma_{\rm at}=\omega Un_{\uparrow}/[\omega -U(1-n_{\uparrow})]
\end{equation}
where $n_{\uparrow}=N_{\uparrow}/N$. This corresponds to the Green's function
\begin{equation}\label{Gat}
G_{\rm at}=(\omega-\Sigma_{\rm at})^{-1}=\frac{n_{\uparrow}}{\omega-U}+\frac{1-n_{\uparrow}}{\omega}
\end{equation}
as obtained by Hubbard~\cite{Hu63}.
Clearly $\Sigma_3(\KK,\omega)$ includes the simple second order diagram which
describes the weak interaction limit as well as sufficient diagrams to describe
the atomic limit. This is why the state  $\vert P_3(\KK)\rangle$ yields a
smooth crossover from the weak coupling "polaron" limit to the strong coupling
limit with "molecule" states forming a lower Hubbard band around $\omega=U
(U<0)$. 

\section{An alternative ansatz}\label{sec:alternative}

Edwards~\cite{Ed90,vdLE91} introduced an ansatz which is equivalent to assuming
that the function $C(Q)$ in the exact wave-function of Eq.~\eqref{psiK} is
a determinant of one-particle orbitals $\phi_s(\kk),s=1...N_{\uparrow}$.
In the real-space (site) representation of the Hubbard model this becomes
\begin{equation}\label{alternative}
\vert \chi(\KK)\rangle =\frac{1}{\surd N}\sum_i {\rm e}^{\ii\KK\cdot\xx_i}c_{i\downarrow}^{\dagger}\prod_s(\sum_j\phi_s
(\xx_j-\xx_i)c_{j\uparrow}^{\dagger})\vert 0\rangle
\end{equation}
where $c_{i\sigma}^{\dagger}$ creates a particle of spin $\sigma$ at the
lattice site $\xx_i$. This ansatz is exact in 1D~\cite{Ed90} but contains
only $NN_{\uparrow}$ variational parameters compared with $\binom{N}{N_{\uparrow}}$
parameters in $\vert \psi(\kk)\rangle$ given by~\eqref{psiK}. This economy
is due to the fact that in 1D with on-site interactions ($\delta$-function
interaction in the gas case) particles entering a collision with wave-vectors
$k_1, k_2$ emerge with the same or interchanged wave-vectors. This is the
basis of the Bethe ansatz. The same is not true in 2D or 3D but, certainly
in 2D, $\vert \chi(\KK)\rangle$ remains a very good approximation in the
Hubbard model~\cite{So94,vdLE91}. For $U>0$ it has been tested by comparison
with accurate results for small clusters~\cite{vdLE91}.

In the gas case the wave-function $\vert \chi(\KK)\rangle$ becomes
\begin{equation}\label{chigas}
\vert \chi_{\rm gas}(\KK)\rangle=(N_{\uparrow}!V)^{-1/2}{\rm e}^{\ii\KK\cdot\xx_0}\rm
det(\phi_s
(\xx_j-\xx_0))
\end{equation}
where $\xx_0$ is the position of the $\downarrow$ spin particle, $\xx_j(j=1...N_{\uparrow})$
are positions of the $\uparrow$ spin particles, $\phi_s (s=1...N_{\uparrow})$
are the orbitals with respect to which the energy of $\vert \chi_{\rm gas}(\KK)\rangle$
is minimised and $V$ is the volume of the gas. The functions $\phi_s(\xx)$
are chosen to be orthonormal. For this gas case the Hamiltonian~\eqref{ham}
may be written as
\begin{equation}\label{hgas}
H_{\rm gas}=-(\hbar^2/2m)(\nabla_0^2 +\Sigma_{{1\leq j\leq N_{\uparrow}}}\nabla_j^2)+g\Sigma_j
\delta(\xx_j-\xx_0),
\end{equation}
with the $\delta$-function regularised in 2D and 3D by the momentum cutoff.
Equations for the orthonormal functions $\phi_s$ are obtained by minimizing
the expectation value $\langle\chi_{\rm gas}(\KK)\vert H_{\rm gas}\vert\ \chi_{\rm gas}(\KK)\rangle$
with respect to variations in these functions. These take the form~\cite{Ed90}
\begin{equation}\label{hartree}
-\nabla^2\phi_s+\rm i\KK\cdot\nabla\phi_s+\sum_r\int\phi_r^*\nabla\phi_s
d^3 r\cdot\nabla\phi_r-\sum_r\int\phi_r^*\nabla\phi_r
d^3 r\cdot\nabla\phi_s+\frac{m}{\hbar^2}[g\delta(\xx)+\lambda_s]\phi_s(\xx)=0
\end{equation}
The volume integrals have been written as three-dimensional but the form
of the equations holds in any number of dimensions. For $\KK=0$ they take
the form of Hartree-Fock equations for a system of particles of mass $m/2$
moving in a short-range potential $g\delta(\xx)$ with interactions $(\mathbf{p}_i\cdot\mathbf{p}_j)/m$,
where $\mathbf{p}_i$ is the momentum of particle i. Equations analogous to~\eqref{hartree} have been obtained for the Hubbard case~\cite{Ed90,vdLE91}
and it is not difficult to show that in the low density limit they reduce
to~\eqref{hartree} with $\hbar^2 /2m=a^2 t, g=Ua^d$ as expected.

We now wish to consider qualitatively the cross-over from the "polaron" to
"molecule" regime from the present viewpoint. In 1D, where the present formalism
is exact, one of the orbitals $\phi_s$ is a localized bound state for any
negative interaction, and the degree of localization grows with increasing
interaction strength. Furthermore as $g$ decreases through 0, from positive
values where there is no bound state, physical properties such as the ground
state energy have been shown to vary continuously with a smooth crossover~\cite{McG66}.
Since any attractive potential also binds in 2D we expect
a bound orbital to develop much as in 1D. However there is a crucial difference
as regards the unbound orbitals. In 1D they are linear combinations of plane
waves with wave-vectors which are not the usual ones corresponding to periodic
boundary conditions, although the complete orbitals do satisfy these conditions.
A consequence of this structure is that there is no quasiparticle weight~\cite{CZ93}. In the 2D Hubbard case Sorella~\cite{So94} has made detailed numerical calculations using the present formalism to show convincingly that
there
is no transition to a state with zero quasiparticle weight for either attractive
or repulsive interaction.

In 3D a bound state will first appear at a critical strength of the attractive
interaction. This situation was considered by Kohn and Majumdar~\cite{KM65}
for a static impurity. In the present context this corresponds to neglecting
the Hartree-Fock terms in~\eqref{hartree} which arise from the motion
of the $\downarrow$ spin particle. Kohn and Majumdar showed that properties
such as the particle density function and ground state energy are smooth
(analytic) functions of the interaction strength, even at the critical value
where a bound state appears. There seems little doubt that this continuity
will still hold in the presence of weak interactions such as the momentum-dependent
ones appearing in~\eqref{hartree}. We therefore expect a continuous
crossover in 1D, 2D and 3D in agreement with the conclusions of section~\ref{sec:exact
wave-function}.

\section{Relation between the two approaches}\label{sec:relation}

In this section we show how approximate wave-functions discussed in section~\ref{sec:exact wave-function} can emerge from the ansatz of section~\ref{sec:alternative}.
The first example we consider is the wave-function $\vert P_3(\KK)\rangle$
which corresponds to~\eqref{psipolaron} with only the first two terms
retained on the right-hand side. This form is obtained from the ansatz~\eqref{alternative}
by taking each orbital $\phi_s$ as a plane wave $(1/\surd N){\rm e}^{\ii\qq\cdot\xx}$
with a small correction, where $\qq$ lies within the FS and the label $s$ may be replaced by $\qq$. Thus
\begin{equation}\label{orbital}
\phi_{\qq}(\xx)=(1/\surd N){\rm e}^{\ii\qq\cdot\xx}+\psi_{\qq}(\xx)
\end{equation}
and~\eqref{alternative} becomes
\begin{equation}\label{newchi}
\vert\chi(\KK)\rangle=\frac{1}{\surd N}\sum_i {\rm e}^{\ii(\KK-\sum\qq)\cdot\xx_i}c_{i\downarrow}^{\dagger}\displaystyle
\prod_{\qq}(c_{\qq\uparrow}^{\dagger}+{\rm e}^{\ii\qq\cdot\xx_i}\sum_j\psi_{\qq}(\xx_j-\xx_i)c_{j\uparrow}^{\dagger})\vert
0\rangle.
\end{equation}
On expanding the product, and retaining only terms up to first order in $\psi_{\qq}$,
we obtain
\begin{equation}\label{nextchi}
\vert\chi(\KK)\rangle=\frac{1}{\surd N}\sum_i {\rm e}^{\ii\KK\cdot\xx_i}c_{i\downarrow}^{\dagger}(1+\sum_{\qq}\rho(\qq) {\rm e}^{\ii\qq\cdot\xx_i}\sum_j\psi_{\qq}(\xx_j-\xx_i)c_{j\uparrow}^{\dagger}c_{\qq\uparrow})\vert\rm FS ( N_{\uparrow})\rangle
\end{equation}
where the factor $\rho(\qq)$ takes values $\pm 1$. Here we have taken $\KK$
relative to the total wave-vector of the FS, as in~\eqref{psipolaron}.
On making a plane-wave expansion of $\psi_{\qq}$,
\begin{equation}\label{psiq}
\rho(\qq)\psi_{\qq}(\xx)=(1/\surd N)\sum_{\kk}\alpha_{\kk\qq}{\rm e}^{\ii\kk\cdot\xx},
\end{equation}
we find
\begin{equation}\label{chifinal}
\vert\chi(\KK)\rangle =( c_{\KK \downarrow}^{\dagger}+\sum_{\kk,\qq}
\alpha_{\kk\qq}c_{\KK -\kk+\qq \downarrow}^{\dagger}c_{\kk\uparrow}^{\dagger}c_{\qq\uparrow})\vert\rm FS ( N_{\uparrow})\rangle
\end{equation}
as required. In general, by expanding the product in~\eqref{newchi} to
order $n$ in $\psi_{\qq}$, we obtain a wave-function of the form $\vert P_{2n+1}(\KK)\rangle$.
However all coefficients in that general form are now expressed in terms
of the comparatively small number of quantities  $\alpha_{\kk\qq}$.

To obtain wave-functions of the form $\vert M_{2n}(\KK)\rangle$ given by~\eqref{M} from the ansatz of Eq.~\eqref{alternative} we take one of the
orbitals $\phi_s(\xx)$ to be a localized function representing a "molecule".
The remaining orbitals are treated as perturbed plane waves, just as before.
Writing the localized orbital as
\begin{equation}\label{philoc}
\phi(\xx)=(1/\surd N)\sum_\kk\xi_{\kk}{\rm e}^{\ii\kk\cdot\xx},
\end{equation}
and using~\eqref{psiq} for the perturbation to the plane waves, we obtain
a first-order expression having the form of $\vert M_4(\KK)\rangle$ defined
by the first two terms on the right-hand side of~\eqref{M}. The coefficients
in the second term are found to be given by $\xi_{\kk\kk_1\qq_1}=\xi_{\kk}\alpha_{\kk_1\qq_1}$.

We have shown that the ansatz $\vert\chi(\KK)\rangle$ given by~\eqref{alternative}
encompasses approximate wave-functions of both the "polaron" and "molecule"
type introduced by other authors. This emphasizes the point made in section~\ref
{sec:exact wave-function} that these two types of state are not physically
distinct and merge continuously into each other.

\section{Conclusion}\label{sec:conclusion}

We consider a system with a single $\downarrow$ spin particle interacting
with a Fermi sea of effectively non-interacting $\uparrow$ spin particles.
It is shown that the transition from "polaron" to "molecule" behaviour is a smooth crossover, not a sharp transition. Consequently the $\downarrow$
spin quasiparticle remains fermionic, with non-zero quasiparticle weight,
even for large interaction strength where it also behaves as part of a molecule.  

This conclusion has been reached for the case when only a single $\downarrow$ spin particle
is involved in interactions. The problem of a finite number $N_{\downarrow}$
of $\downarrow$ spin particles, in particular the balanced spin case $ N_{\downarrow}=N_{\uparrow}$,
is the much more complex one of the full attractive Hubbard model. The considerations
of this paper throw no light on the nature of the transition between the
Bardeen-Cooper-Schrieffer (BCS) and the Bose-Einstein-condensation (BEC)
limits of the superconducting state in this model. 

\section*{Acknowledgement}
I was stimulated to write this paper by the "Condensed Matter Physics in
the City 2013" programme and I am grateful to the organisers for inviting
me to participate. 

\section*{References}

\begin{thebibliography}{10}

\bibitem{Hu63}
Hubbard J 1963 {\em Proc. R. Soc.} A {\bf 276} 238
\bibitem{McG65}
McGuire J~B 1965 {\em J. Math. Phys.} {\bf 6} 432
\bibitem{McG66}
McGuire J~B 1966 {\em J. Math. Phys.} {\bf 7} 123
\bibitem{LW69}
Lieb E and Wu FY 1968 {\em Phys. Rev. Lett.} {\bf 20} 1445
\bibitem{Ed90}
Edwards D~M 1990 {\em Prog. Theor. Phys. Suppl.} {\bf 101} 453
\bibitem{CZ93}
Castella H and Zotos X 1993 {\em  Phys. Rev.} B {\bf 47} 16186
\bibitem{Na66}
Nagaoka Y 1966 {\em  Phys. Rev.} {\bf 147} 392
\bibitem{Ro69}
Roth L 1969 {\em  Phys. Rev.} {\bf 186} 428
\bibitem{vdLE91}
von der Linden W and Edwards D M 1991 {\em J. Phys. Condens. Matter} {\bf 3} 4917
\bibitem{Wu96}
Wurth P, Uhrig G S and M{\"u}ller-Hartmann E 1996 {\em Ann. Phys. (Leipzig)}
{\bf 5} 148
\bibitem{An90} 
Anderson P~W 1990  {\em Phys. Rev. Lett.} {\bf64} 1839; {\bf65} 2306
\bibitem{So94}
Sorella S 1994 {\em  Phys. Rev.} B {\bf 49} 12373
\bibitem{CM10}
Chevy F and Mora C 2010 {\em Rep. Prog. Phys.} {\bf 73} 112401
\bibitem{PSa08}
Prokof'ev N and Svistunov B 2008 {\em  Phys. Rev.} B {\bf 77} 020408(R)
\bibitem{PSb08}
Prokof'ev N and Svistunov B 2008 {\em  Phys. Rev.} B {\bf 77} 125101
\bibitem{PDZ09}
Punk M, Dumitrescu P~T and Zwerger W 2009  {\em  Phys. Rev.} A {\bf 80} 053605
\bibitem{BM10}
Bruun GM and Massignan P 2010 {\em Phys. Rev. Lett.} {\bf105} 020403
\bibitem{TC12}
Trefzger C and Castin Y 2012 {\em  Phys. Rev.} A {\bf 85} 053612
\bibitem{Pa11}
Parish M~M 2011 {\em  Phys. Rev.} A {\bf 83} 051603(R)
\bibitem{PL13}
Parish M~M and Levinsen J  2013 {\em  Phys. Rev.} A {\bf 87} 033616
\bibitem{ZBP11}
Z{\"o}llner S, Bruun G M and Pethick C J  2011 {\em  Phys. Rev.} A {\bf 83} 021603(R)
\bibitem{CG08}
Combescot R and Giraud S 2008 {\em Phys. Rev. Lett.} {\bf101} 050404
\bibitem{C06}
Chevy F 2006  {\em  Phys. Rev.} A {\bf 74} 063628
\bibitem{CRLC07}
Combesco R, Recati A, Lobo C and Chevy F 2007  {\em Phys. Rev. Lett.} {\bf98} 180402
\bibitem{Ta75}
Tan B W 1975 {\em PhD thesis} London
\bibitem{KM65}
Kohn W and Majumdar C 1965 {\em Phys. Rev.} {\bf 138} A 1617

\end{thebibliography}

\end{document}